\documentclass[aps,prl,twocolumn,superscriptaddress,longbibliography]{revtex4-2}
\usepackage{amsmath}
\usepackage{amssymb}
\usepackage{graphicx}
\usepackage[
  bookmarks=false,
  pdfpagemode=UseNone,
  colorlinks=true,
  linkcolor=blue,
  citecolor=blue,
  urlcolor=blue
]{hyperref}

\begin{document}

\title{Fluctuation-based evidence for number--phase dynamics \\ in a frustrated orbital superfluid}

\author{Rui-Lang Zeng}
\thanks{These authors have contributed equally to this work.}
\affiliation{State Key Laboratory of Quantum Functional Materials, Department of Physics and Institute for Quantum Science and Engineering, and Guangdong Basic Research Center of Excellence for Quantum Science, Southern University of Science and Technology, Shenzhen 518055, China}
\author{Zi-Yao Zhang}
\thanks{These authors have contributed equally to this work.}
\affiliation{State Key Laboratory of Quantum Functional Materials, Department of Physics and Institute for Quantum Science and Engineering, and Guangdong Basic Research Center of Excellence for Quantum Science, Southern University of Science and Technology, Shenzhen 518055, China}
\author{Ling-Na Wu}
\thanks{These authors have contributed equally to this work.}
\affiliation{Center for Theoretical Physics and School of Physics and Optoelectronic Engineering, Hainan University, Haikou, Hainan 570228, China}
\author{Cong-Jie Zhang}
\affiliation{State Key Laboratory of Quantum Functional Materials, Department of Physics and Institute for Quantum Science and Engineering, and Guangdong Basic Research Center of Excellence for Quantum Science, Southern University of Science and Technology, Shenzhen 518055, China}
\author{Da-Gang Xia}
\affiliation{State Key Laboratory of Quantum Functional Materials, Department of Physics and Institute for Quantum Science and Engineering, and Guangdong Basic Research Center of Excellence for Quantum Science, Southern University of Science and Technology, Shenzhen 518055, China}
\author{Andreas Hemmerich}
\affiliation{Institute of Quantum Physics, University of Hamburg, Luruper Chaussee 149, 22761 Hamburg, Germany}
\author{Xiao-Qiong Wang}
\email{wangxq@sustech.edu.cn}
\affiliation{State Key Laboratory of Quantum Functional Materials, Department of Physics and Institute for Quantum Science and Engineering, and Guangdong Basic Research Center of Excellence for Quantum Science, Southern University of Science and Technology, Shenzhen 518055, China}
\author{Zhi-Fang Xu}
\email{xuzf@sustech.edu.cn}
\affiliation{State Key Laboratory of Quantum Functional Materials, Department of Physics and Institute for Quantum Science and Engineering, and Guangdong Basic Research Center of Excellence for Quantum Science, Southern University of Science and Technology, Shenzhen 518055, China}

\begin{abstract}
Frustrated quantum matter can host intertwined orders rooted in symmetry-related low-energy landscapes, yet static order parameters alone do not reveal how fluctuations are organized among competing configurations. Here we measure mode-resolved shot-to-shot population fluctuations in a $p$-orbital triangular-lattice superfluid with a tunable bias among three valleys. We observe a bias-tuned evolution from enhanced, anticorrelated fluctuations of two minority valleys toward strong confinement of relative-population fluctuations in a selected two-valley stripe phase. The dominant fluctuation structure is captured by an effective canonical model that includes interactions among the condensed modes, supporting a quasi-equilibrium description of the coherent three-valley condensate. Together, the data and model reveal a quantum--thermal regime shaped by pair-tunneling-induced number--phase dynamics, in which relative-phase scrambling softens effective barriers in the minority-valley regime, while phase rigidity gives rise to macroscopic harmonic confinement in the stripe phase. Our results establish mode-resolved fluctuation measurements as a probe of hidden number--phase back-action in frustrated quantum fluids.
\end{abstract}

\maketitle

Complex quantum materials often conceal their organizational principles by exhibiting several intertwined broken symmetries~\cite{KivelsonReview}. Across cuprate superconductors~\cite{Agterberg2020,Hamidian2016,Comin2015}, twisted moir\'e superlattices~\cite{Cao2018,Andrei2020,Kennes2021,Cao2021,Jiang2019} and Kagome metals~\cite{Jiang2021,Chen2021,Nie2022,Neupert2022}, charge order, nematicity and pair-density-wave order can arise as closely connected instabilities involving symmetry-related wavevectors or orientations. Static order alone does not reveal how these competing components fluctuate and correlate within the underlying low-energy landscape~\cite{Fernandes2019,Hwangbo2024}. Their joint fluctuation statistics are difficult to access in solid-state systems, where experiments typically identify a selected broken-symmetry state rather than repeatedly sample equivalent preparations.

Ultracold atoms provide a complementary route for exploring this statistical scenario~\cite{Bloch2008,Bloch2012}. When bosons condense in multiple valleys, the order parameter is specified by complex amplitudes for the occupied modes, forming a discrete multicomponent order parameter. Repeated preparation together with single-shot momentum-space imaging makes their joint population fluctuations and correlations experimentally accessible~\cite{Altman2004,Folling2005,Schweigler2017}. This enables fluctuation-resolved studies of collective order beyond its average static structure. Recent observations of Brownian motion of a Bose--Einstein condensate further illustrate that fluctuation measurements can reveal collective dynamics hidden by ensemble averaging~\cite{Wang2025}.

Optical lattices offer controllable routes to multivalley condensates, either through periodic lattice driving~\cite{Struck2013,Parker2013} or through occupation of higher orbital bands~\cite{Wirth2011,Kock2016,Wang2021,Wang2023}. Two-valley condensates have provided paradigmatic settings for $Z_2$ symmetry breaking, order selection and nonequilibrium domain dynamics~\cite{Parker2013,Clark2016}. A $p$-orbital triangular lattice realizes a minimal three-valley extension of this setting. Quantum stripe order, anticipated for interacting $p$-orbital bosons~\cite{Wu2006}, was previously established in this platform~\cite{Wang2023}, while the hidden number--phase dynamics of the three-valley condensate remained unresolved. Unlike density interactions that only shape the static landscape, phase-sensitive inter-valley pair tunneling redistributes atoms and couples a valley-pair imbalance to its relative phase~\cite{Hemmerich2019}. Even in a weakly interacting condensate, frustration and pair tunneling can organize collective fluctuations that are invisible to a static mean-field order parameter. Characterizing their quantum and thermal structure therefore requires access to the joint fluctuation statistics.

Here, we address this task by using mode-resolved shot-to-shot fluctuations to probe the long-time statistical state of the orbital superfluid. By tuning the relative energy landscape of the three valleys, we compare the measured fluctuation and correlation observables with an effective canonical model that includes interactions among the condensed modes. The data are naturally described by a quasi-equilibrium distribution of the coherent three-valley condensate, rather than by a simple frozen mixture of independently prepared domains. This comparison separates a frustrated minority-valley regime with enhanced minority-valley fluctuations and strong anticorrelations from a selected two-valley stripe phase whose residual relative fluctuations are described by a macroscopic harmonic oscillator.
Together, these two limits reveal how inter-valley pair tunneling couples valley-population imbalance to its conjugate relative phase, producing relative-phase scrambling and barrier softening on one side, and phase rigidity with harmonic confinement on the other.

% ==============================================================================
% Figure 1: (RMS Fluctuations)
% ==============================================================================
\begin{figure}[t]
\centering
\includegraphics[width=\linewidth]{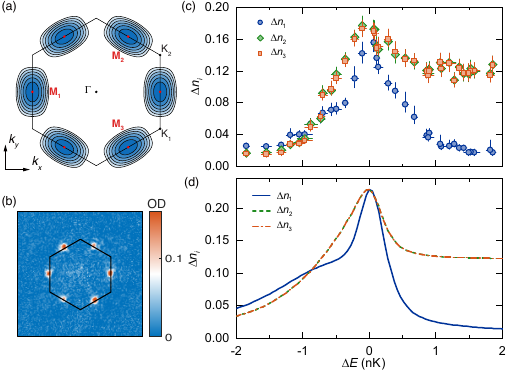}
\caption{\label{fig:fig1} 
(a) Schematic $p$-orbital band structure of the triangular lattice, with three minima at the $M$ points labeled $M_1$, $M_2$ and $M_3$. The bias $\Delta E$ tunes the energy of $M_1$ relative to $M_2$ and $M_3$; negative (positive) $\Delta E$ lowers (raises) $M_1$.
(b) Representative time-of-flight image after dissipative preparation, showing sharp Bragg peaks at the three $M$ valleys. Hexagons in (a) and (b) mark the first Brillouin zone.
(c) Measured root-mean-square fluctuations of the normalized valley populations, $\Delta n_i$, as a function of $\Delta E$. Horizontal error bars denote the uncertainty in $\Delta E$; vertical error bars are obtained from bootstrap resampling.
(d) Canonical three-mode calculation of $\Delta n_i$ for $T/N=0.04~\mathrm{nK}$, capturing the main bias-dependent fluctuation hierarchy.
}
\end{figure}

\textit{Fluctuations in a three-valley condensate}---%
Our experiment begins with an ultracold gas of $^{87}\mathrm{Rb}$ atoms loaded into the $p$-orbital band of a two-dimensional triangular optical lattice. The single-particle band has three minima at the $M$ points of the Brillouin zone, labeled $M_i$ with energies $E_i$ $(i=1,2,3)$, as illustrated in Fig.~\ref{fig:fig1}(a). After a fixed dissipative evolution time of $130.1~\mathrm{ms}$, an orbital superfluid has formed. Momentum-space images obtained after time of flight show sharp Bragg peaks at the three $M$ valleys, indicating that the signal is dominated by a coherent condensate in the three-valley manifold (Fig.~\ref{fig:fig1}(b)). We then tune the relative valley energies through the bias $\Delta E=E_1-(E_2+E_3)/2$, with $E_2\simeq E_3$, as described previously~\cite{Wang2023}, and measure how the condensate fluctuations evolve across this bias-tuned energy landscape.

For each value of $\Delta E$, we repeat the preparation and extract the populations $N_i$ condensed at the three $M_i$ valleys. We define normalized valley populations $n_i=N_i/N$, where $N=\sum_iN_i$, and analyze the shot-to-shot fluctuations $\delta n_i=n_i-\langle n_i\rangle$, with root-mean-square amplitudes $\Delta n_i=\sqrt{\langle(\delta n_i)^2\rangle}$. As shown in Fig.~\ref{fig:fig1}(c), the measured fluctuations are redistributed non-monotonically with bias. In the large negative-bias regime, where $M_1$ hosts the dominant condensate (cf. Fig.~2(b) in Ref.~\cite{Wang2023}), $\Delta n_1$ is slightly larger than the fluctuations of the two depleted valleys. Approaching three-valley degeneracy from the negative side, the hierarchy reverses and the minority-valley fluctuations $\Delta n_2$ and $\Delta n_3$ exceed $\Delta n_1$, showing that the depleted valleys form an active fluctuation channel. For positive bias, $\Delta n_1$ is suppressed as $M_1$ is depleted, while the $M_2$ and $M_3$ fluctuations level off in the selected two-valley stripe regime. This hierarchy is not fixed simply by the mean valley populations or the single-particle bias, suggesting an interaction-reshaped fluctuation structure within the coherent three-valley manifold.

To interpret this fluctuation map, we use an effective three-mode description of the condensed $M$ valleys. The sharp Bragg peaks observed in the momentum-space images (Fig.~\ref{fig:fig1}(b)) justify treating each valley as a single macroscopic mode. The corresponding interacting Hamiltonian is
\begin{align}
\hat{H} &= \Delta E\,\hat{b}^{\dagger}_1\hat{b}_1
+ \frac{U_1}{2} \sum_j \hat{b}_j^\dagger \hat{b}_j^\dagger \hat{b}_j \hat{b}_j
+ 2U_2\sum_{i<j}\hat{b}_i^\dagger \hat{b}_j^\dagger \hat{b}_j \hat{b}_i \nonumber\\
&\quad
+ \frac{U_2}{2} \sum_{i<j}\left(\hat{b}_i^\dagger \hat{b}_i^\dagger \hat{b}_j \hat{b}_j+\mathrm{h.c.}\right),
\end{align}
where $\hat{b}_j$ annihilates an atom in the $M_j$ valley, while $U_1$ and $U_2$ denote the intra- and inter-valley interaction matrix elements, respectively. The first term describes the tunable valley bias, the next two terms are density interactions, and the last term is inter-valley pair tunneling. The single-particle and density-interaction terms conserve the individual valley populations, whereas pair tunneling provides the coherent channel for population redistribution among the three $M$ valleys and links valley-population imbalance to relative-phase dynamics. Dissipative preparation removes excess energy and enables relaxation toward the measured distribution of this coherent three-valley condensate.

The measured shot-to-shot distributions are compared with a canonical ensemble of this condensed three-mode system. Because the absolute in situ temperature and coherent atom number are difficult to determine independently from time-of-flight images, we consider a single effective thermodynamic ratio, $T/N$, from the overall fluctuation amplitude. For fixed interaction scales $U_1N$ and $U_2N$, this ratio primarily controls the canonical distribution. Using $U_1N=3.32~\mathrm{nK}$ and $U_2N=1.82~\mathrm{nK}$, we find that $T/N=0.04~\mathrm{nK}$ captures the main structure of the measured fluctuation trajectory; the corresponding calculation is shown in Fig.~\ref{fig:fig1}(d). This agreement supports an effective quasi-equilibrium description within the coherent three-valley condensate, in a regime that remains coherent enough for pair-tunneling-induced number--phase dynamics and thermal enough for canonical fluctuation sampling.

The comparison in Figs.~\ref{fig:fig1}(c) and \ref{fig:fig1}(d) therefore identifies an anomalous negative-bias minority-valley response, a susceptibility maximum near three-valley degeneracy, and a positive-bias stripe regime with confined relative fluctuations. These regimes provide the experimental basis for the pair-tunneling-induced conjugate dynamics analyzed below.

% ==============================================================================
% Figure 2: (C23)
% ==============================================================================
\begin{figure}[t]
\centering
\includegraphics[width=\linewidth]{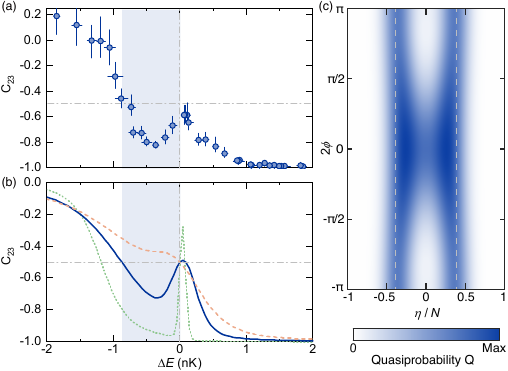}
\caption{\label{fig:fig2}
(a) Correlation coefficient $C_{23}$ between the two minority valleys as a function of $\Delta E$. The vertical and horizontal dash-dotted lines mark $\Delta E=0$ and the symmetric three-valley reference value $C_{23}=-0.5$, respectively. The shaded region indicates the anomalous regime with $C_{23}<-0.5$.
(b) Canonical three-mode calculations of $C_{23}$ for different effective thermodynamic ratios $T/N$. The thick blue curve corresponds to the experimentally constrained value $T/N=0.04~\mathrm{nK}$, while the dotted green and dashed orange curves correspond to $T/N=0.01~\mathrm{nK}$ and $0.08~\mathrm{nK}$, respectively.
(c) Husimi $Q$ distribution from the canonical three-mode calculation at $\Delta E=-0.35~\mathrm{nK}$ and $T/N=0.04~\mathrm{nK}$, near the minimum of the calculated dip in (b). The distribution shows two competing minority-valley configurations in $\eta/N$ and a broad distribution along $2\phi$. The vertical dashed lines mark $\eta/N=\pm\langle n_2+n_3\rangle$.
}
\end{figure}

\textit{Anomalous minority-valley anticorrelations}---%
We first focus on the negative-bias regime, $\Delta E<0$, where the single-particle bias favors the $M_1$ valley. The enhanced root-mean-square fluctuations of the two minority valleys in Fig.~\ref{fig:fig1}(c) show that these depleted modes form an active fluctuation channel, but they do not by themselves reveal whether $M_2$ and $M_3$ fluctuate together or compete with one another. To distinguish these possibilities, we evaluate the normalized correlation coefficient $C_{23}=\langle\delta n_2\delta n_3\rangle/(\Delta n_2\Delta n_3)$ between the two minority valleys.

The reference value of $C_{23}$ follows from normalization, $\sum_i n_i=1$. For equal three-valley variances, this constraint gives $C_{23}=-0.5$. As shown in Fig.~\ref{fig:fig2}(a), the measured $C_{23}$ approaches this value near three-valley degeneracy, but develops a deeper dip, close to $-0.8$, at intermediate negative bias. This dip is nontrivial because $M_2$ and $M_3$ are not an isolated two-mode system with a fixed combined population: the total minority fraction can still fluctuate through redistribution with the favored $M_1$ valley, which would push $C_{23}$ upward rather than below the balanced reference. A value below $-0.5$ therefore signals that internal-imbalance fluctuations dominate over combined-population fluctuations. To make this explicit, decomposing the minority-valley fluctuations into the total minority fraction $s=n_2+n_3$ and the relative imbalance $d=n_2-n_3$ gives, for $\Delta n_2\simeq\Delta n_3$, $C_{23}\simeq[(\Delta s)^2-(\Delta d)^2]/[(\Delta s)^2+(\Delta d)^2]$. Thus $C_{23}<-0.5$ implies $(\Delta d)^2>3(\Delta s)^2$, showing that the negative-bias dip identifies binary competition between the two minority valleys rather than fluctuations of their total population.

The canonical three-mode calculation reproduces this non-monotonic correlation structure, as shown in Fig.~\ref{fig:fig2}(b). Varying $T/N$ changes both the depth and width of the $C_{23}$ minimum: lower effective temperature drives the system toward the ideal binary-competition limit $C_{23}\rightarrow -1$, whereas higher temperature washes out the binary structure and pushes $C_{23}$ back above the $-0.5$ threshold. The experimentally constrained ratio $T/N=0.04~\mathrm{nK}$ captures the observed anomalous dip, indicating equilibrium-like sampling of two competing minority-valley configurations within the interacting condensate.

This equilibrium-like sampling is nontrivial when compared with a rigid phase-locked semiclassical landscape. Writing $\hat b_j\simeq\sqrt{N_j}e^{-i\theta_j}$, the macroscopically occupied $M_1$ condensate acts as a phase reservoir: the dominant $M_1$--$M_2$ and $M_1$--$M_3$ pair-tunneling channels lock the minority-valley phases to $M_1$, with $\theta_2-\theta_1=\pm\pi/2$ and $\theta_3-\theta_1=\pm\pi/2$. Their mutual relative phase is then fixed to $\phi=\theta_2-\theta_3=0$ or $\pi$, so the direct $M_2$--$M_3$ pair-tunneling term penalizes balanced occupation and forms a classical barrier between the two imbalanced minority-valley configurations. Taking $N_{\mathrm{act}}=N_2+N_3\simeq N/2$ gives the representative upper-bound estimate $E_{\mathrm{b}}^{\mathrm{cl}}\simeq N^2(3U_2-U_1)/16\simeq3.4\,k_{\mathrm{B}}T$~\cite{suppl}. Such a barrier would not forbid equilibration in principle, but it would make ordinary thermal sampling kinetically unfavorable on the finite experimental timescale if the phase-locked landscape remained rigid.

The inferred sampling of both minority configurations therefore raises a kinetic question: how can an equilibrium-like distribution emerge within the finite evolution time if the relevant landscape remains rigidly phase locked with a barrier of this scale? In the pair-tunneling model, the valley-population imbalance $\eta=N_2-N_3$ is conjugate to the relative phase $\phi$, with $[\hat{\phi},\hat{\eta}]=2i$. When the imbalance becomes comparable to $N_{\mathrm{act}}$, it acts as a large conjugate momentum for the relative phase. This gives a phase-evolution scale $E_{\mathrm{phase}}\simeq |U_1-3U_2|N_{\mathrm{act}}$~\cite{suppl}. Taking $N_{\mathrm{act}}\simeq N/2$ gives $E_{\mathrm{phase}}\simeq1.1~\mathrm{nK}$, corresponding to a characteristic phase-spreading time $\tau_{\phi}=h/E_{\mathrm{phase}}\simeq45~\mathrm{ms}$, shorter than the $130.1~\mathrm{ms}$ evolution time used for the fluctuation measurements.

Because the phase velocity is tied to the fluctuating imbalance, this dynamics broadens the relative-phase distribution. A leading-cumulant estimate gives
$\langle\cos(2\phi)\rangle\simeq \cos(2\langle\phi\rangle)e^{-2\langle(\delta\phi)^2\rangle}$.
Near the phase-locked configurations, $\langle\phi\rangle=0$ or $\pi$, so an $\mathcal{O}(1)$ growth of the phase variance already exponentially suppresses the phase-sensitive pair-tunneling penalty, without requiring a uniform phase distribution over the full $2\pi$ circle. The effective barrier is thereby reduced to $E_{\mathrm{b}}^{\mathrm{flat}}\simeq N^2(2U_2-U_1)/16\simeq0.5\,k_{\mathrm{B}}T$. Relative-phase broadening therefore renormalizes the barrier to the thermal scale, making both minority-valley configurations accessible to residual thermal fluctuations.

The same canonical calculation visualizes this phase-broadened landscape through the Husimi $Q$ distribution. As shown in Fig.~\ref{fig:fig2}(c), near the calculated anticorrelation dip the distribution has weight near two competing configurations in $\eta/N$, while remaining broad along the conjugate pair-tunneling phase $2\phi$. Although this is not a direct phase measurement, the phase-space structure shows that the canonical state consistent with the measured anticorrelations is phase broadened and barrier softened, rather than rigidly phase locked. It therefore gives model-supported evidence for relative-phase scrambling as one mechanism making both minority-valley configurations statistically accessible.

%==============================================================================
% Figure 3:  (Total Variance)
%==============================================================================
\begin{figure}[t]
\centering
\includegraphics[width=5.6cm]{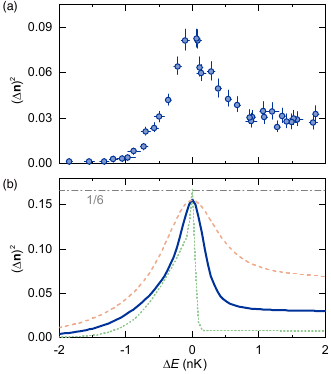}
\caption{\label{fig:fig3}
(a) Total fluctuation metric $(\Delta \mathbf{n})^2$ as a function of $\Delta E$, showing a pronounced peak near the symmetric point $\Delta E\simeq0$.
(b) Canonical three-mode calculations of $(\Delta \mathbf{n})^2$ for different effective thermodynamic ratios $T/N$. The thick solid, dotted and dashed curves correspond to $T/N=0.04~\mathrm{nK}$, $0.01~\mathrm{nK}$ and $0.08~\mathrm{nK}$, respectively. The horizontal dash-dotted line marks the zero-temperature discrete-manifold limit $(\Delta\mathbf n)^2=1/6$.
}
\end{figure}

\textit{Maximal susceptibility at three-valley degeneracy}---%
We next examine the nearly symmetric regime, $\Delta E\simeq0$, where the three valleys are closest to degeneracy. To characterize the collective valley response, we use the rotationally invariant fluctuation metric $(\Delta\mathbf{n})^2\equiv\sum_{i=1}^{3}(\Delta n_i)^2$. As shown in Fig.~\ref{fig:fig3}(a), this quantity exhibits a pronounced peak near $\Delta E=0$, identifying the point where the free-energy landscape is softest and the condensate is most susceptible to redistribution within the three-valley manifold.

In the zero-temperature three-mode limit, the discrete stripe manifold gives the population-fluctuation reference value $(\Delta\mathbf{n})^2=1/6$. At finite temperature, the canonical model broadens this discrete-manifold response into a susceptibility fan around $\Delta E=0$, as shown in Fig.~\ref{fig:fig3}(b): lower effective temperature sharpens the peak, whereas higher temperature broadens it over a wider bias range.

The measured peak is lower than the ideal three-mode prediction at the experimentally constrained thermodynamic ratio. This reduced amplitude may reflect the sensitivity of the nearly degenerate point to effects beyond a homogeneous equilibrium three-mode description, including dilution by thermally populated modes around the $M$ points, spatial averaging over regions with different local parameters, and slow finite-time relaxation in the exceptionally soft landscape. Thus, the model captures the peak position and overall structure, while the amplitude is reduced by effects beyond the idealized limit.

% ==============================================================================
% Figure 4:  (Phase Rigidity)
% ==============================================================================
\begin{figure}[t]
\centering
\includegraphics[width=\linewidth]{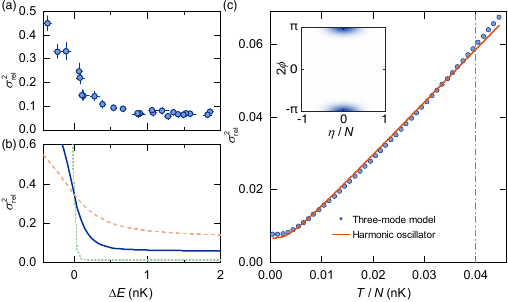}
\caption{\label{fig:fig4}
(a) Measured normalized relative variance $\sigma_{\mathrm{rel}}^2$ as a function of $\Delta E$. Its rapid decay and weakly varying plateau indicate confinement of relative-population fluctuations in the selected two-valley stripe phase.
(b) Canonical three-mode calculations of $\sigma_{\mathrm{rel}}^2$ for different effective thermodynamic ratios $T/N$. The thick blue, dotted green and dashed orange curves correspond to $T/N=0.04~\mathrm{nK}$, $0.01~\mathrm{nK}$ and $0.08~\mathrm{nK}$, respectively.
(c) Temperature dependence of $\sigma_{\mathrm{rel}}^2$ at $\Delta E=3~\mathrm{nK}$. Blue circles show the canonical three-mode results, and the orange curve shows the harmonic-oscillator fit derived from the phase-rigid expansion. The gray dash-dotted line marks the experimentally constrained ratio $T/N=0.04~\mathrm{nK}$. Inset: Husimi $Q$ distribution at the same bias and thermodynamic ratio.
}
\end{figure}

\textit{Phase rigidity in the selected stripe phase}---%
Finally, we turn to the positive-bias regime, $\Delta E>0$, where the bias suppresses the $M_1$ population and selects the two-valley stripe phase formed by $M_2$ and $M_3$. To quantify the residual fluctuations within this phase, we evaluate the normalized relative variance
$\sigma_{\mathrm{rel}}^2\equiv\langle(\delta n_2-\delta n_3)^2\rangle/\langle n_2+n_3\rangle^2$.
As shown in Fig.~\ref{fig:fig4}(a), $\sigma_{\mathrm{rel}}^2$ decreases rapidly with increasing $\Delta E$ and then approaches a weakly varying plateau. The canonical three-mode calculation at the experimentally constrained thermodynamic ratio captures this suppression of relative imbalance fluctuations, as shown in Fig.~\ref{fig:fig4}(b). Lower effective temperatures drive the relative fluctuations toward the zero-point limit, whereas higher temperatures maintain larger thermal imbalance fluctuations over a wider bias range.

The physical origin of this confinement is relative-phase rigidity. A static mean-field picture fixes the condensate in an equal-weight superposition of $M_2$ and $M_3$, but it does not describe the residual conjugate fluctuations around this state. In the selected stripe phase, the $M_2$--$M_3$ pair-tunneling interaction generates a Josephson-like phase potential that locks the relative phase at $\phi=\pm\pi/2$. Since the valley-population imbalance $\eta$ is conjugate to $\phi$, this phase rigidity confines, but does not eliminate, the residual imbalance fluctuations. Expanding the pair-tunneling Hamiltonian around the phase-locked minimum therefore maps the stripe phase onto a macroscopic harmonic oscillator.

This oscillator description gives a unified quantum--thermal form for the confined fluctuations. The unnormalized variance $\langle(\delta\hat{\eta})^2\rangle$ scales linearly with the active two-valley population $N_{\mathrm{act}}$~\cite{suppl}, so that the measured quantity $\sigma_{\mathrm{rel}}^2\simeq\langle(\delta\hat{\eta})^2\rangle/N_{\mathrm{act}}^2$ decreases as $1/N_{\mathrm{act}}$ in the phase-rigid regime. This scaling is characteristic of a collective coherent oscillator, rather than an incoherent mixture of macroscopically separated population states. The temperature dependence in Fig.~\ref{fig:fig4}(c) shows how the harmonic-oscillator picture connects thermal equipartition to the quantum zero-point limit. The Husimi $Q$ distribution in the inset provides the corresponding phase-space visualization: the distribution is localized near $\eta=0$ and $2\phi=\pm\pi$, illustrating the phase-rigid stripe condition and the confinement of the conjugate population imbalance.

In summary, mode-resolved fluctuations in this system are not merely residual noise around an ordered state. When combined with the canonical three-mode model, these measurements reveal how competing valley configurations are statistically organized within a frustrated condensate manifold. The anomalous minority-valley anticorrelations and the suppressed relative-population fluctuations in the selected stripe phase are complementary manifestations of pair-tunneling-induced number--phase back-action. The model supports relative-phase scrambling and barrier softening in the minority-valley regime, and phase rigidity with harmonic confinement of the conjugate population imbalance in the stripe phase. More generally, mode-resolved fluctuation measurements provide access to collective correlations and dynamics beyond static mean-field order parameters in frustrated quantum fluids.

\begin{acknowledgments}
We acknowledge Guang-Quan Luo for contributions at the early stage of this work and thank Zi-Xiang Li for helpful discussions. This work was supported by the National Key R\&D Program of China (Grant No.~2022YFA1404103) and the National Natural Science Foundation of China (Grant Nos.~12274196, 92476101, 12522412, and 12304289).
\end{acknowledgments}


% Supplemental Material appended for the combined arXiv version.
\clearpage
\onecolumngrid
\setcounter{equation}{0}
\setcounter{figure}{0}
\setcounter{table}{0}
\renewcommand{\theequation}{S\arabic{equation}}
\renewcommand{\thefigure}{S\arabic{figure}}
\renewcommand{\thetable}{S\arabic{table}}
\renewcommand{\theHequation}{S\arabic{equation}}
\renewcommand{\theHfigure}{S\arabic{figure}}
\renewcommand{\theHtable}{S\arabic{table}}

\begin{center}
{\Large \textbf{Supplemental Material}}
\end{center}

\section*{Supplemental Note 1: Effective three-mode model and parameter estimates}

This note summarizes the effective three-mode model, parameter estimates, canonical calculation and phase-space representation used for the calculations in the main text. The interaction scales are obtained by projecting the contact interaction onto the calibrated three-valley Bloch modes, including the vertical confinement and an effective filling estimate. Once the macroscopic scales $U_1N$ and $U_2N$ are fixed, the fixed-$N$ canonical ensemble is controlled primarily by the effective thermodynamic ratio $T/N$, which is constrained by the global bias-dependent fluctuation data. We therefore document the lattice calibration, overlap integrals, filling estimate and parameter uncertainty underlying the theoretical curves. We also describe the Husimi $Q$ phase-space representation used to visualize the canonical density matrix in the conjugate variables of valley-population imbalance and relative phase.

\subsection*{S1A. Effective three-mode Hamiltonian}

For reference, we reproduce the effective three-mode Hamiltonian used in the main text. The relevant condensed manifold is spanned by the three second-band minima at the $M$ points of the triangular optical lattice, labeled $M_i$ with single-particle energies $E_i$ $(i=1,2,3)$. We denote the corresponding annihilation operators by $\hat b_i$. After subtracting an irrelevant common single-particle energy, the projected Hamiltonian takes the form
\begin{align}
\hat H
&=
\Delta E\,\hat b_1^\dagger \hat b_1
+\frac{U_1}{2}\sum_{i=1}^{3}
\hat b_i^\dagger \hat b_i^\dagger \hat b_i \hat b_i
+2U_2\sum_{i<j}
\hat b_i^\dagger \hat b_j^\dagger \hat b_j \hat b_i
+\frac{U_2}{2}\sum_{i<j}
\left(
\hat b_i^\dagger\hat b_i^\dagger \hat b_j\hat b_j
+\mathrm{h.c.}
\right).
\label{eq:S1_H}
\end{align}
Here $\Delta E=E_1-(E_2+E_3)/2$ is the experimentally controlled valley bias, with $E_2\simeq E_3$ maintained in the experiment. The coefficient $U_1$ describes intra-valley repulsion, while $U_2$ sets the inter-valley density interaction and pair-tunneling scales within the three-valley manifold. The pair-tunneling term is central to the number--phase dynamics discussed in the main text.

The total atom number in the projected manifold,
\begin{equation}
\hat N=\sum_{i=1}^{3}\hat b_i^\dagger\hat b_i,
\label{eq:S1_N}
\end{equation}
is fixed in the canonical calculation. The many-body basis is therefore given by Fock states $|N_1,N_2,N_3\rangle$ satisfying $N_1+N_2+N_3=N$.

\subsection*{S1B. Optical-lattice calibration and band calculation}

The energy bias $\Delta E$ is obtained from a single-particle band calculation using the calibrated final triangular-lattice configuration. The two-dimensional optical potential is written as
\begin{equation}
V(\mathbf{r})=-V_A\left[3+2\sum_{j=1}^{3}\alpha_j\cos\left(\mathbf{b}_j\cdot\mathbf{r}\right)\right],
\label{eq:S1_lattice}
\end{equation}
where $V_A$ is the triangular-lattice depth and $\alpha_j$ parameterize the calibrated relative amplitudes of the three interference terms. We use the beam wavevectors $\mathbf{k}_1=(-\sqrt{3}/2,1/2)k_L$, $\mathbf{k}_2=(\sqrt{3}/2,1/2)k_L$ and $\mathbf{k}_3=(0,-1)k_L$, with $k_L=2\pi/\lambda$, and define $\mathbf{b}_1=\mathbf{k}_1-\mathbf{k}_2$, $\mathbf{b}_2=\mathbf{k}_2-\mathbf{k}_3$ and $\mathbf{b}_3=\mathbf{k}_3-\mathbf{k}_1$.

The lattice depths are calibrated by amplitude-modulation spectroscopy. Although the final triangular lattice is formed by interference of three lattice beams, the calibration is performed pairwise: for each measurement, only two beams are turned on, forming a one-dimensional optical lattice. The modulation frequency is scanned and the remaining atom number is measured to obtain an atom-loss spectrum. At each modulation frequency, the measurement is repeated three times, giving the mean atom number and its standard deviation. The averaged resonance profile is fitted with a Gaussian function to determine the resonance minimum, and the fitting uncertainty gives the uncertainty of the measured excitation frequency. By comparing this resonance frequency with the calculated $\Gamma$-point $s$-to-$d$ band gap of the corresponding one-dimensional lattice, we obtain the lattice depth and its uncertainty for that beam pair. Repeating this procedure for the three beam pairs gives the calibrated depths of the three interference terms. For each bias setting, these calibrated parameters are then used in a plane-wave band calculation to determine the second-band energies at the three $M$ valleys and hence $\Delta E$. Propagating the lattice-depth uncertainties through this calculation gives the horizontal error bars in the main figures.

For the interaction-parameter estimates, the Bloch functions are evaluated at the symmetric final triangular lattice, with $\alpha_j=1$ and $V_A=0.71E_R$ for $\lambda=1064~\mathrm{nm}$, where $E_R=h^2/(2m\lambda^2)$. The vertical mode is approximated by a harmonic oscillator using the measured optical-dipole-trap frequency $\omega_z=2\pi\times69.5~\mathrm{Hz}$. We neglect the weak dependence of $U_1$ and $U_2$ on the small intensity imbalance used to tune the bias.

\subsection*{S1C. Interaction matrix elements}

The interaction parameters are obtained by projecting the contact interaction onto the three $M$-valley Bloch modes. We write the three-dimensional mode function as
$\Phi_i(\mathbf{r}_\perp,z)=\psi_i(\mathbf{r}_\perp)\chi(z)$, where $\psi_i$ is the second-band Bloch function at $M_i$ calculated from Eq.~\eqref{eq:S1_lattice}, and $\chi(z)$ is the ground-state wavefunction of the vertical harmonic confinement. The contact interaction strength is $g=4\pi\hbar^2a_s/m$, and the vertical overlap factor is $I_z=\int dz|\chi(z)|^4=1/(\sqrt{2\pi}a_z)$, with $a_z=\sqrt{\hbar/(m\omega_z)}$.

The two interaction scales in Eq.~\eqref{eq:S1_H} are then written as
\begin{equation}
U_1=gI_z\mathcal I_1,\qquad
U_2=gI_z\mathcal I_2,
\label{eq:S1_U1U2}
\end{equation}
where the two-dimensional overlap integrals are
\begin{equation}
\mathcal I_1=\int d^2\mathbf{r}_\perp\,|\psi_i(\mathbf{r}_\perp)|^4,\qquad
\mathcal I_2=\int d^2\mathbf{r}_\perp\,
\psi_i^*(\mathbf{r}_\perp)\psi_i^*(\mathbf{r}_\perp)
\psi_j(\mathbf{r}_\perp)\psi_j(\mathbf{r}_\perp),
\label{eq:S1_overlap}
\end{equation}
with $i\neq j$. At the time-reversal-invariant $M$ points, the Bloch functions can be chosen real, so $\mathcal I_2$ also equals the density overlap $\int d^2\mathbf{r}_\perp\,|\psi_i|^2|\psi_j|^2$. This is why the same coefficient $U_2$ enters both the inter-valley density and pair-tunneling terms in Eq.~\eqref{eq:S1_H}, while $U_1$ denotes the intra-valley interaction.

\subsection*{S1D. Filling estimate and thermodynamic scaling}

Because the Bloch functions are normalized over the occupied two-dimensional lattice, the overlap integrals $\mathcal I_1$ and $\mathcal I_2$ scale approximately as $1/N_u$, where $N_u$ is the number of occupied unit cells. The macroscopic interaction scales are therefore controlled by the effective filling $\nu=N/N_u$. For the representative filling $\nu\simeq8$ used in the main calculation, the microscopic overlap estimates give $U_1N=3.318~\mathrm{nK}$ and $U_2N=1.824~\mathrm{nK}$.

The canonical calculation is performed at fixed total atom number $N$ in the projected three-valley manifold. Thermal averages are evaluated using
\begin{equation}
\hat{\rho}_T=
\frac{e^{-\hat H/k_{\mathrm B}T}}{Z},
\qquad
Z=\mathrm{Tr}_{N}\left(e^{-\hat H/k_{\mathrm B}T}\right),
\label{eq:S1_rhoT}
\end{equation}
where $\mathrm{Tr}_{N}$ denotes the trace over the fixed-$N$ Fock space. For fixed $U_1N$ and $U_2N$, the normalized population distributions and fluctuation observables are controlled primarily by the effective thermodynamic ratio $T/N$.

The value $T/N=0.04~\mathrm{nK}$ is constrained by the global comparison with the measured bias-dependent fluctuation trajectory, rather than by fitting a single data point. Smaller or larger values of $T/N$ give visibly different thermal sharpening or broadening of the calculated fluctuation features, as illustrated by the reference curves in the main figures. The corresponding temperature scale should be viewed as an effective thermodynamic scale inferred from the long-time coherent three-valley fluctuation data. It is expected to be comparable to the low-energy temperature scale of the atomic cloud, although it is not obtained from an independent thermometry measurement.

\subsection*{S1E. Parameter uncertainty and robustness}

The main uncertainties in the absolute interaction scales arise from the effective filling, the vertical confinement entering $I_z$, and the use of an effective three-mode description for the trapped cloud. These uncertainties mainly affect the overall scale of $U_1N$ and $U_2N$. The ratio $U_1/U_2$ is more robust, because both coefficients are obtained from the same Bloch functions and share the same vertical overlap factor. We also neglect the weak dependence of the Bloch functions on the small intensity imbalance used to tune $\Delta E$.

Changing the overall interaction scale primarily rescales the bias range over which the theoretical features appear. This effect cannot be fully compensated by retuning $T/N$, because the feature positions, the hierarchy of $\Delta n_i$, and the correlation structure provide independent constraints. We therefore use a single representative parameter set in the main comparison, constrained by the microscopic overlap estimate and by the global consistency with the measured fluctuation curves. Within reasonable parameter variations, the enhanced minority-valley fluctuations for $\Delta E<0$, the susceptibility peak near three-valley degeneracy, and the suppression of relative fluctuations for $\Delta E>0$ remain qualitatively robust.

\subsection*{S1F. Husimi \(Q\) phase-space representation}

To visualize the phase-space structure of the three-mode equilibrium state, we evaluate the Husimi \(Q\) distribution \(Q(\eta/N,2\phi)\) from the canonical density matrix \(\hat{\rho}_T\) used to calculate the fluctuation observables in the main text. This analysis is not an additional fit to the data, but a phase-space representation of the same finite-temperature three-mode model.

We use fixed-\(N\) three-mode coherent states,
\begin{equation}
|\Omega\rangle=\frac{1}{\sqrt{N!}}\left(\sum_{j=1}^{3}\sqrt{f_j}\,e^{-i\theta_j}\hat b_j^\dagger\right)^N|0\rangle ,
\label{eq:S1_coherent_operator}
\end{equation}
where \(f_j\geq0\), \(\sum_j f_j=1\), and \(\theta_j\) are the mode phases. This convention matches the semiclassical parametrization $\hat b_j\simeq\sqrt{N_j}e^{-i\theta_j}$ used throughout the analysis.
In the Fock basis \(|\boldsymbol{\ell}\rangle=|\ell_1,\ell_2,\ell_3\rangle\), with \(\ell_1+\ell_2+\ell_3=N\), this state becomes
\begin{equation}
|\Omega\rangle=\sum_{\boldsymbol{\ell}}\left(\frac{N!}{\ell_1!\ell_2!\ell_3!}\right)^{1/2}f_1^{\ell_1/2}f_2^{\ell_2/2}f_3^{\ell_3/2}e^{-i(\ell_1\theta_1+\ell_2\theta_2+\ell_3\theta_3)}|\boldsymbol{\ell}\rangle .
\label{eq:S1_coherent_fock}
\end{equation}
Here \(f_j\) are coherent-state population fractions and should not be confused with the normalized populations \(n_j\) used for the experimental fluctuation observables.

For the \(M_2\)--\(M_3\) phase-space representation used in Fig.~2(c) and the inset of Fig.~4(c), we use \(\eta=N_2-N_3\) and \(\phi=\theta_2-\theta_3\), and take \(M_1\) as the phase reference by setting \(\theta_1=0\). The remaining phases are parameterized as
\begin{equation}
\theta_2=\Theta+\frac{\phi}{2},\qquad \theta_3=\Theta-\frac{\phi}{2},
\label{eq:S1_phase_param}
\end{equation}
where \(\Theta=(\theta_2+\theta_3)/2\) is an unobserved common phase. The phase factor then separates as
\begin{equation}
\ell_2\theta_2+\ell_3\theta_3=\Theta(\ell_2+\ell_3)+\frac{2\phi}{4}(\ell_2-\ell_3).
\label{eq:S1_phase_decomposition}
\end{equation}

For a fixed value of \(\eta/N\), the remaining population fraction \(f_1\) determines
\begin{equation}
f_2=\frac{1-f_1+\eta/N}{2},\qquad f_3=\frac{1-f_1-\eta/N}{2},\qquad 0\leq f_1\leq1-|\eta|/N .
\label{eq:S1_fractions_Q}
\end{equation}
We define the \(\Theta\)-independent coefficient
\begin{equation}
C_{\boldsymbol{\ell}}(\eta/N,f_1,2\phi)=\left(\frac{N!}{\ell_1!\ell_2!\ell_3!}\right)^{1/2}f_1^{\ell_1/2}f_2^{\ell_2/2}f_3^{\ell_3/2}e^{-i(2\phi)(\ell_2-\ell_3)/4}.
\label{eq:S1_Cell_Q}
\end{equation}
The coherent state can then be written as
\begin{equation}
|\Omega(\Theta)\rangle=\sum_{\boldsymbol{\ell}}C_{\boldsymbol{\ell}}e^{-i\Theta(\ell_2+\ell_3)}|\boldsymbol{\ell}\rangle .
\label{eq:S1_Omega_Theta}
\end{equation}

The common phase \(\Theta\) is integrated out. Writing \((\rho_T)_{\boldsymbol{\ell}^{\prime},\boldsymbol{\ell}}=\langle\boldsymbol{\ell}^{\prime}|\hat{\rho}_T|\boldsymbol{\ell}\rangle\), the phase average gives
\begin{equation}
\frac{1}{2\pi}\int_0^{2\pi}d\Theta\,\langle\Omega(\Theta)|\hat{\rho}_T|\Omega(\Theta)\rangle=\sum_{\boldsymbol{\ell}^{\prime},\boldsymbol{\ell}}C_{\boldsymbol{\ell}^{\prime}}^{*}(\rho_T)_{\boldsymbol{\ell}^{\prime},\boldsymbol{\ell}}C_{\boldsymbol{\ell}}\,\delta_{\ell^{\prime}_1,\ell_1}.
\label{eq:S1_Theta_trace}
\end{equation}
Here \(C_{\boldsymbol{\ell}}\) denotes \(C_{\boldsymbol{\ell}}(\eta/N,f_1,2\phi)\). The Kronecker delta follows from \(\ell^{\prime}_2+\ell^{\prime}_3=\ell_2+\ell_3\), which is equivalent to \(\ell^{\prime}_1=\ell_1\) at fixed total particle number. We therefore define \((\tilde{\rho}_T)_{\boldsymbol{\ell}^{\prime},\boldsymbol{\ell}}=(\rho_T)_{\boldsymbol{\ell}^{\prime},\boldsymbol{\ell}}\delta_{\ell^{\prime}_1,\ell_1}\).

The Husimi \(Q\) distribution shown in the main figures is thus given by
\begin{equation}
Q(\eta/N,2\phi)=\mathcal{N}_Q\int_0^{1-|\eta|/N}df_1\,\sum_{\boldsymbol{\ell}^{\prime},\boldsymbol{\ell}}C_{\boldsymbol{\ell}^{\prime}}^{*}(\eta/N,f_1,2\phi)(\tilde{\rho}_T)_{\boldsymbol{\ell}^{\prime},\boldsymbol{\ell}}C_{\boldsymbol{\ell}}(\eta/N,f_1,2\phi),
\label{eq:S1_Q_function}
\end{equation}
where \(\mathcal{N}_Q\) normalizes the distribution. In the numerical implementation, the multinomial factors in Eq.~\eqref{eq:S1_Cell_Q} are evaluated in logarithmic form, and Eq.~\eqref{eq:S1_Q_function} is evaluated over the \(2\phi\) grid by matrix operations after constructing \(\tilde{\rho}_T\).

The Husimi \(Q\) distribution provides a phase-space visualization of the same canonical density matrix used for the fluctuation observables. In the negative-bias regime, Fig.~2(c) shows two competing minority-valley configurations in \(\eta/N\), together with a broad distribution along the conjugate pair-tunneling phase \(2\phi\). In the positive-bias regime, the inset of Fig.~4(c) is concentrated near \(\eta=0\) and \(2\phi=\pm\pi\), visualizing the phase-rigid \(M_2\)--\(M_3\) stripe condition and the confinement of the conjugate population imbalance. Since \(2\phi\) is periodic, \(2\phi=\pi\) and \(-\pi\) represent the same phase-locked stripe condition.

\section*{Supplemental Note 2: Negative-bias anticorrelations and phase-scrambling mechanism}

This note supports the negative-bias analysis in the main text. We first relate the anomalous condition $C_{23}<-0.5$ to internal imbalance fluctuations of the two minority valleys. We then estimate the phase-locked classical barrier and show how number--phase conjugacy provides a route to relative-phase scrambling, which softens this barrier and allows equilibrium-like sampling of the two competing minority-valley configurations.

\subsection*{S2A. Minority-valley correlations and binary competition}

We characterize the competition between the two minority valleys by the normalized correlation coefficient $C_{23}=\langle\delta n_2\delta n_3\rangle/(\Delta n_2\Delta n_3)$. Since the normalized populations satisfy $n_1+n_2+n_3=1$ in each shot, one has $\delta n_1=-(\delta n_2+\delta n_3)$ and therefore
\begin{equation}
(\Delta n_1)^2=(\Delta n_2)^2+(\Delta n_3)^2+2\langle\delta n_2\delta n_3\rangle .
\label{eq:S2_variance_relation}
\end{equation}
For a symmetric three-valley fluctuation pattern with $\Delta n_1=\Delta n_2=\Delta n_3$, this relation gives $C_{23}=-0.5$. For approximately symmetric minority valleys, $\Delta n_2\simeq\Delta n_3$, the same relation gives $(\Delta n_1)^2\simeq2(\Delta n_2)^2(1+C_{23})$. Thus $C_{23}<-0.5$ is equivalent to the anomalous hierarchy $\Delta n_1<\Delta n_2\simeq\Delta n_3$ observed in the negative-bias window.

To distinguish total minority-population fluctuations from internal minority-valley imbalance fluctuations, we introduce $s=n_2+n_3$ and $d=n_2-n_3$. Using $n_2=(s+d)/2$ and $n_3=(s-d)/2$, one obtains, for $\Delta n_2\simeq\Delta n_3$,
\begin{equation}
C_{23}\simeq\frac{(\Delta s)^2-(\Delta d)^2}{(\Delta s)^2+(\Delta d)^2}.
\label{eq:S2_C23_sd}
\end{equation}
Therefore $C_{23}<-0.5$ corresponds to $(\Delta d)^2>3(\Delta s)^2$. The anomalous dip below the symmetric reference thus identifies fluctuations dominated by the internal imbalance of the minority-valley pair, rather than by fluctuations of the total minority population.

As a useful zero-temperature reference, exact diagonalization shows that the negative-bias low-energy states are organized by two imbalanced minority-valley configurations. A representative schematic form is
\begin{equation}
|\Psi\rangle\simeq\sum_k c_k\left(|N-k,k,0\rangle+|N-k,0,k\rangle\right),
\label{eq:S2_branch_state}
\end{equation}
where the two components correspond to finite minority population in either $M_2$ or $M_3$. The precise coefficients and low-energy level structure are not used below; Eq.~\eqref{eq:S2_branch_state} only serves as a limiting reference for the binary minority-valley structure inferred from the fluctuation correlations. The correlation analysis by itself identifies this binary competition; the following sections address why the competing configurations can be sampled in the finite-time experiment rather than remaining kinetically trapped in a rigid phase-locked landscape.

\subsection*{S2B. Phase-locked classical barrier}

To obtain a classical reference landscape, we write the condensed valley modes as $\hat b_j\simeq\sqrt{N_j}e^{-i\theta_j}$. The interaction energy associated with Eq.~\eqref{eq:S1_H} is
\begin{equation}
E_{\mathrm{int}}\simeq
\frac{U_1}{2}\sum_{j=1}^{3}N_j^2
+2U_2\sum_{i<j}N_iN_j
+U_2\sum_{i<j}N_iN_j\cos\left[2(\theta_i-\theta_j)\right].
\label{eq:S2_Eint}
\end{equation}
For $\Delta E<0$, the macroscopically occupied $M_1$ valley acts as a phase reference. The pair-tunneling terms involving $M_1$ are minimized when the minority-valley phases are locked relative to $M_1$, with $\theta_2-\theta_1=\pm\pi/2$ and $\theta_3-\theta_1=\pm\pi/2$. This phase anchoring constrains the mutual relative phase $\phi=\theta_2-\theta_3$ to $0$ or $\pi$, so that the direct $M_2$--$M_3$ pair-tunneling term contributes $+U_2N_2N_3$. For fixed total minority population, this term penalizes simultaneous occupation of $M_2$ and $M_3$ and forms a classical barrier between the two imbalanced minority-valley configurations.

With this phase locking, the part of the interaction energy controlling the competition between $M_2$ and $M_3$ is
\begin{align}
F_{23}^{\mathrm{cl}}
&\simeq \frac{U_1}{2}(N_2^2+N_3^2)+3U_2N_2N_3 
= \frac{U_1+3U_2}{4}N_{\mathrm{act}}^2+\frac{U_1-3U_2}{4}\eta^2 ,
\label{eq:S2_Fcl_eta}
\end{align}
where $\eta=N_2-N_3$ and $N_{\mathrm{act}}=N_2+N_3$. For fixed $N_{\mathrm{act}}$, the first term is independent of $\eta$. Since $3U_2>U_1$ for the interaction parameters used here, the symmetric point $\eta=0$ is the top of the classical barrier in the phase-locked landscape, while the two imbalanced minority-valley configurations correspond to $\eta\simeq\pm N_{\mathrm{act}}$.

The corresponding classical barrier height is therefore
\begin{equation}
E_{\mathrm b}^{\mathrm{cl}}
=F_{23}^{\mathrm{cl}}(\eta=0)-F_{23}^{\mathrm{cl}}(\eta=\pm N_{\mathrm{act}})
\simeq \frac{(3U_2-U_1)N_{\mathrm{act}}^2}{4}.
\label{eq:S2_Ebcl_Nact}
\end{equation}
Taking $N_{\mathrm{act}}\simeq N/2$ as a representative upper-bound estimate in the bistable window gives
\begin{equation}
E_{\mathrm b}^{\mathrm{cl}}\simeq \frac{N^2(3U_2-U_1)}{16}.
\label{eq:S2_Ebcl}
\end{equation}
The measured active minority population near the anticorrelation dip is smaller than $N/2$, so this value should be viewed as an upper-bound scale for the phase-locked barrier. With the parameter set and effective thermodynamic ratio of Supplemental Note~1, this gives $E_{\mathrm b}^{\mathrm{cl}}/k_{\mathrm B}T\simeq3.37$. In such a phase-locked landscape, ordinary thermal sampling across this barrier would be strongly suppressed, favoring kinetic trapping near one imbalanced minority-valley configuration. As an order-of-magnitude microscopic reference, the relevant $p$-orbital tunneling amplitudes in the triangular optical lattice are of order $10~\mathrm{nK}$~[27], corresponding to $\tau_0\sim h/[k_{\mathrm B}(10~\mathrm{nK})]\sim5~\mathrm{ms}$. Combining this microscopic reference with the representative upper-bound barrier gives an illustrative single-crossing scale $\tau_{\mathrm{cross}}^{\mathrm{cl}}\sim\tau_0\exp(E_{\mathrm b}^{\mathrm{cl}}/k_{\mathrm B}T)\sim150~\mathrm{ms}$, comparable to the $130.1~\mathrm{ms}$ evolution time. In this representative phase-locked estimate, the experimental window would therefore contain only an order-unity number of thermally activated crossings, whereas robust equilibrium-like mixing would require repeated exploration of both minority-valley configurations. This estimate is intended only as an order-of-magnitude kinetic benchmark. The compatibility of the measured correlation statistics with the canonical three-mode calculation therefore calls for an additional mechanism that softens this classical barrier on the experimental timescale.

\subsection*{S2C. Conjugate phase uncertainty and phase-spreading scale}

The phase-locked classical barrier assumes that the relative phase remains well defined. In the quantum three-mode model, however, the minority-valley imbalance $\hat{\eta}=\hat N_2-\hat N_3$ and the relative phase $\hat{\phi}=\hat\theta_2-\hat\theta_3$ are conjugate variables. In the usual number--phase approximation,
\begin{equation}
[\hat{\phi},\hat{\eta}]=2i,\qquad \Delta\eta\,\Delta\phi\geq1 .
\label{eq:S2_uncertainty}
\end{equation}
This number--phase uncertainty provides a natural route to relative-phase scrambling through the growth of the phase variance $\langle(\delta\hat{\phi})^2\rangle$. When, within either imbalanced configuration, the population concentrates in one minority valley, the other minority mode becomes weakly occupied and its number fluctuations are reduced; for approximately fixed $N_{\mathrm{act}}$, this narrows the imbalance uncertainty $\Delta\eta$, so the conjugate relative phase is no longer sharply defined. The resulting broad relative-phase distribution suppresses the phase-sensitive average $\langle\cos(2\hat{\phi})\rangle$.

To discuss the stability of the phase-locked reference, one must retain the phase locking imposed by the macroscopic $M_1$ reservoir. The reservoir-locked configurations have $\theta_2-\theta_1=\pm\pi/2$ and $\theta_3-\theta_1=\pm\pi/2$, so that the $M_2$--$M_3$ relative phase is locked at $\phi=0$ or $\pi$. These two locked configurations have the same quadratic stability. For compactness, we expand around the $\phi=0$ configuration and write
\begin{equation}
\theta_2-\theta_1=\frac{\pi}{2}+\zeta+\frac{\phi}{2},\qquad
\theta_3-\theta_1=\frac{\pi}{2}+\zeta-\frac{\phi}{2},
\label{eq:S2_phase_param}
\end{equation}
where $\zeta$ is the common phase offset relative to $M_1$, and $\phi=\theta_2-\theta_3$ is the local deviation from the chosen locked configuration. The locked configuration at $\phi=\pi$ is obtained by shifting $\phi$ by $\pi$ and gives the same quadratic result.

Expressed in these variables, the full interaction energy near a reservoir-locked configuration is
\begin{align}
E_{\mathrm{int}}
=&\,\frac{U_1}{2}N_1^2+\frac{U_1}{4}\left(N_{\mathrm{act}}^2+\eta^2\right)
+2U_2N_1N_{\mathrm{act}}+\frac{U_2}{2}\left(N_{\mathrm{act}}^2-\eta^2\right) \nonumber\\
&-U_2N_1\left[N_{\mathrm{act}}\cos(2\zeta)\cos\phi-\eta\sin(2\zeta)\sin\phi\right]
+\frac{U_2}{4}\left(N_{\mathrm{act}}^2-\eta^2\right)\cos(2\phi).
\label{eq:S2_full_phase_energy}
\end{align}
This expression keeps both the $M_1$--minority phase-locking terms and the direct $M_2$--$M_3$ pair-tunneling term.

Expanding Eq.~\eqref{eq:S2_full_phase_energy} around $\zeta=0$, $\phi=0$ and $\eta=0$ gives, to quadratic order,
\begin{equation}
\delta E_{\mathrm{int}}
\simeq
2U_2N_1N_{\mathrm{act}}\zeta^2
+K\phi^2-G\eta^2,
\qquad
K=\frac{U_2}{2}N_{\mathrm{act}}\left(N_1-N_{\mathrm{act}}\right),
\qquad
G=\frac{3U_2-U_1}{4}>0 .
\label{eq:S2_local_instability}
\end{equation}
Equation~\eqref{eq:S2_local_instability} shows the self-limiting character of the phase-locked classical reference. In the negative-bias regime considered here, $N_1>N_{\mathrm{act}}$, so $K>0$ and the relative phase remains locally locked. However, this stiffness is reduced as the active minority population grows, because the direct $M_2$--$M_3$ pair-tunneling term contributes with the opposite curvature. In contrast, the negative term $-G\eta^2$ means that the balanced minority configuration $\eta=0$ is unstable along the imbalance direction. The phase-locked reference is therefore a saddle point of the local energy landscape, rather than a stable harmonic minimum.

Within this local quadratic approximation, the dynamics near the saddle point is described by an inverted oscillator,
\begin{equation}
\hat H_{\mathrm{eff}}\simeq K\hat{\phi}^{2}-G\hat{\eta}^{2}.
\label{eq:S2_inverted_oscillator}
\end{equation}
Here $\hat{\phi}$ denotes the local deviation from either locked value, $\phi=0$ or $\pi$. Together with Eq.~\eqref{eq:S2_uncertainty}, Eq.~\eqref{eq:S2_inverted_oscillator} gives the growth rate
\begin{equation}
\Gamma_{\mathrm{inv}}=\frac{4\sqrt{KG}}{\hbar}.
\label{eq:S2_Gamma_inv}
\end{equation}

For an initially uncorrelated fluctuation state satisfying $\langle\delta\hat{\eta}_0\delta\hat{\phi}_0+\delta\hat{\phi}_0\delta\hat{\eta}_0\rangle=0$, with $\delta\hat{\phi}_0=\hat{\phi}(0)-\langle\hat{\phi}(0)\rangle$ and $\delta\hat{\eta}_0=\hat{\eta}(0)-\langle\hat{\eta}(0)\rangle$, the relative-phase variance grows as
\begin{equation}
\langle(\delta\hat{\phi})^2(t)\rangle
\simeq
\frac{e^{2\Gamma_{\mathrm{inv}}t}}{4}
\left[
\langle(\delta\hat{\phi}_0)^2\rangle
+\frac{G}{K}\langle(\delta\hat{\eta}_0)^2\rangle
\right],
\label{eq:S2_phi_growth}
\end{equation}
once the unstable growth becomes appreciable. The imbalance variance is amplified in the same conjugate dynamics. Thus an initially phase-locked state is intrinsically prone to relative-phase spreading: small number or phase fluctuations are amplified, driving the system away from a rigid phase-locked classical trajectory.

The quantity relevant for barrier softening is therefore the growth of the phase variance $\langle(\delta\hat{\phi})^2\rangle$ itself. For a locally Gaussian phase distribution, or equivalently at the level of the leading cumulant expansion, the phase-sensitive factor is suppressed as
\begin{equation}
\langle\cos(2\hat{\phi})\rangle
\simeq
\cos(2\langle\hat{\phi}\rangle)
\exp[-2\langle(\delta\hat{\phi})^2\rangle].
\label{eq:S2_Debye_Waller}
\end{equation}
Thus an $O(1)$ increase of the relative-phase variance is already sufficient to exponentially reduce the magnitude of the phase-sensitive pair-tunneling term, without requiring a strictly uniform phase distribution over the full $2\pi$ circle. This Debye--Waller-type renormalization connects the inverted-oscillator growth in Eq.~\eqref{eq:S2_phi_growth} to the softened effective landscape discussed below. The local growth analysis is not used as a quantitative fit, because $K$ depends on the transient populations during preparation; its role is to show why a state initially locked to the $M_1$ reservoir naturally tends toward phase scrambling rather than stable phase rigidity.

For the experimentally relevant scale estimate, we use the corresponding equation of motion,
\begin{equation}
\hbar\frac{d\hat{\phi}}{dt}
\simeq
\left[
U_1-2U_2-U_2\cos(2\hat{\phi})
\right]\hat{\eta}.
\label{eq:S2_dphi}
\end{equation}
Near either phase-locked reference, the phase distribution is centered at $\phi=0$ or $\pi$, so the operator factor $\cos(2\hat{\phi})$ is approximated by unity at the level of this energy-scale estimate. Equation~\eqref{eq:S2_dphi} then reduces to $\hbar d\hat{\phi}/dt\simeq(U_1-3U_2)\hat{\eta}$, giving
\begin{equation}
E_{\mathrm{phase}}\simeq |U_1-3U_2|\,|\langle\hat{\eta}\rangle| .
\label{eq:S2_Ephase}
\end{equation}
For strongly imbalanced minority-valley configurations, $|\langle\hat{\eta}\rangle|$ is bounded by the active minority population. Using the same upper-bound scale, $\langle\hat N_{\mathrm{act}}\rangle\simeq N/2$, gives $E_{\mathrm{phase}}\simeq1.08~\mathrm{nK}$ and a characteristic phase-spreading time $\tau_{\phi}=h/E_{\mathrm{phase}}\simeq44.6~\mathrm{ms}$. This quoted time corresponds to a full $2\pi$ phase-winding scale; an order-unity broadening of the relative phase, sufficient to suppress the phase-sensitive factor, occurs on the shorter scale set by $\hbar/E_{\mathrm{phase}}$. Together with the exponential suppression in Eq.~\eqref{eq:S2_Debye_Waller}, this estimate supports a picture in which the phase-sensitive pair-tunneling term can be substantially reduced within the $130.1~\mathrm{ms}$ experimental window, although the relative phase itself is not directly measured.

\subsection*{S2D. Barrier softening and equilibrium-like sampling}

Relative-phase scrambling softens the barrier by suppressing the phase-sensitive pair-tunneling term. As shown by the Debye--Waller-type renormalization in Eq.~\eqref{eq:S2_Debye_Waller}, growth of the relative-phase variance exponentially reduces the magnitude of $\langle\cos(2\hat{\phi})\rangle$. In the limiting case where this phase-sensitive contribution is strongly suppressed, the effective minority-valley landscape becomes
\begin{align}
F_{23}^{\mathrm{flat}}
&\simeq \frac{U_1}{2}(N_2^2+N_3^2)+2U_2N_2N_3 
= \frac{U_1+2U_2}{4}N_{\mathrm{act}}^2+\frac{U_1-2U_2}{4}\eta^2 .
\label{eq:S2_Fflat_eta}
\end{align}
For fixed $N_{\mathrm{act}}$, the first term is independent of $\eta$. Compared with the phase-locked result in Eq.~\eqref{eq:S2_Fcl_eta}, the phase-sensitive contribution responsible for the large classical curvature is effectively removed. Since $U_1\simeq2U_2$ for the parameters used here, the residual curvature is much smaller than that of the phase-locked landscape.

Because $2U_2>U_1$, the residual barrier height scales as
\[
E_{\mathrm b}^{\mathrm{flat}}
=F_{23}^{\mathrm{flat}}(\eta=0)-F_{23}^{\mathrm{flat}}(\eta=\pm N_{\mathrm{act}})
\simeq \frac{(2U_2-U_1)N_{\mathrm{act}}^2}{4}.
\]
Using the same representative upper-bound estimate $N_{\mathrm{act}}\simeq N/2$ gives
\begin{equation}
E_{\mathrm b}^{\mathrm{flat}}\simeq \frac{N^2(2U_2-U_1)}{16}.
\label{eq:S2_Ebflat}
\end{equation}
With the parameter set of Supplemental Note~1, this yields $E_{\mathrm b}^{\mathrm{flat}}/k_{\mathrm B}T\simeq0.52$, much smaller than the phase-locked estimate $E_{\mathrm b}^{\mathrm{cl}}/k_{\mathrm B}T\simeq3.37$. A smaller active minority population would reduce both barrier estimates by the same factor $N_{\mathrm{act}}^2$, leaving their ratio, $E_{\mathrm b}^{\mathrm{flat}}/E_{\mathrm b}^{\mathrm{cl}}=(2U_2-U_1)/(3U_2-U_1)$, unchanged. Using the same microscopic reference, the representative upper-bound softened barrier gives an illustrative single-crossing scale $\tau_{\mathrm{cross}}^{\mathrm{flat}}\sim\tau_0\exp(E_{\mathrm b}^{\mathrm{flat}}/k_{\mathrm B}T)\sim8~\mathrm{ms}$, more than an order of magnitude shorter than the phase-locked benchmark. This contrast illustrates how barrier softening can substantially accelerate inter-configuration sampling.

The negative-bias regime therefore illustrates the quantum--thermal balance emphasized in the main text. The condensate remains coherent enough for the phase-sensitive pair-tunneling channel to remain active, while number--phase conjugacy allows relative-phase scrambling to soften the barrier to a scale where residual thermal fluctuations can sample the two competing minority-valley configurations. This provides a model-supported mechanism for equilibrium-like sampling of the configurations associated with the strong negative $C_{23}$, without requiring a direct measurement of the relative phase.

\subsection*{S2E. Finite window of anomalous anticorrelations}

The anomalous regime with $C_{23}<-0.5$ appears only within a finite negative-bias window. Near $\Delta E\to0^{-}$, the restoration of three-valley symmetry drives the correlation back toward the symmetric reference value $C_{23}=-0.5$. At more negative bias, the increasing single-particle gap depletes the minority valleys and weakens the nonlinear pair-tunneling competition between the two imbalanced minority-valley configurations. The system then crosses back toward a conventional fluctuation regime. The finite extent of the dip therefore marks the window in which binary minority-valley competition and barrier softening are both relevant, bounded by symmetry restoration near $\Delta E=0$ on one side and minority-valley depletion at larger negative bias on the other.

\section*{Supplemental Note 3: Collective susceptibility near three-valley degeneracy}

This note supports the analysis of the fluctuation peak near $\Delta E\simeq0$. We clarify why the total valley-population variance is a collective susceptibility of the three-valley manifold, derive the zero-temperature discrete-manifold limit, and discuss how modes beyond the discrete condensate approximation can reduce the observed peak height.

\subsection*{S3A. Total variance as a collective valley susceptibility}

The fluctuation metric used in the main text is
\begin{equation}
(\Delta\mathbf n)^2=\sum_{i=1}^{3}(\Delta n_i)^2 .
\label{eq:S3_total_variance}
\end{equation}
Because the normalized populations satisfy $\sum_i n_i=1$, this quantity measures fluctuations within the two-dimensional population simplex of the three valleys. It is invariant under permutation of the valley labels and therefore characterizes the total valley-population fluctuation, independent of which population configuration is selected in a given shot.

The connection to susceptibility follows from the response to small valley-energy offsets. We perturb the three-mode energy by
$\delta H=N\sum_j\delta E_j n_j=\sum_j\delta E_j N_j$, where only relative offsets are physically relevant. In the canonical ensemble,
\begin{equation}
\chi_{ij}
\equiv
-\frac{\partial\langle n_i\rangle}{\partial\delta E_j}
=
\frac{N}{k_{\mathrm B}T}
\langle\delta n_i\delta n_j\rangle .
\label{eq:S3_susceptibility}
\end{equation}
Thus the covariance matrix of the valley populations is proportional to the susceptibility matrix for redistributing atoms among the valleys. The scalar $(\Delta\mathbf n)^2$ is the trace of the covariance matrix, and therefore measures the total collective susceptibility of the three-valley manifold. Near $\Delta E\simeq0$, where the valleys are closest to degeneracy, this susceptibility is maximal because the condensate can be most easily redistributed within the three-valley manifold.

\subsection*{S3B. Zero-temperature discrete-manifold limit}

At zero bias, the zero-temperature three-mode model approaches a discrete stripe manifold. In mean-field language, this manifold contains three population-distinct choices of the occupied valley pair, and each pair has two time-reversed relative-phase choices with phase difference $\pm\pi/2$. A representative stripe state involving valleys $M_i$ and $M_j$ is
\begin{equation}
|\psi_{ij}^{\pm}\rangle=\frac{1}{\sqrt{N!}}\left(\frac{\hat b_i^\dagger\pm \mathrm{i}\hat b_j^\dagger}{\sqrt{2}}\right)^N|0\rangle .
\label{eq:S3_stripe_state}
\end{equation}
In this coherent two-valley condensate, each atom occupies $M_i$ or $M_j$ with probability $1/2$. The atom number in each occupied valley therefore follows a binomial distribution, giving $\langle N_i\rangle=\langle N_j\rangle=N/2$ and $(\Delta N_i)^2=(\Delta N_j)^2=N/4$, while the third valley is unoccupied.

For population observables, the two phase choices of a given valley pair have identical number statistics. Equal sampling over the three population-distinct stripe configurations gives $\langle N_i\rangle=N/3$ for each valley and
\begin{equation}
\langle N_i^2\rangle=\frac{2}{3}\left[\left(\frac{N}{2}\right)^2+\frac{N}{4}\right].
\label{eq:S3_Ni2}
\end{equation}
Thus,
\begin{equation}
(\Delta\mathbf n)^2=\sum_{i=1}^{3}\frac{\langle N_i^2\rangle-\langle N_i\rangle^2}{N^2}
=\frac{1}{6}+\frac{1}{2N}.
\label{eq:S3_limit}
\end{equation}
In the large-$N$ comparison used in the main text, this becomes the zero-temperature discrete-manifold limit $(\Delta\mathbf n)^2=1/6$.

\subsection*{S3C. Peak height and limitations of the discrete three-mode approximation}

At finite temperature, the three-mode calculation broadens the discrete-manifold response into a susceptibility peak around $\Delta E=0$. Lower effective temperatures sharpen the peak toward the zero-temperature population-fluctuation reference value $(\Delta\mathbf n)^2=1/6$, while higher temperatures broaden the response over a wider bias range.

The measured peak height is lower than the ideal three-mode prediction evaluated at the same experimentally constrained thermodynamic ratio. This reduced amplitude likely reflects several limitations of a homogeneous equilibrium discrete three-mode description near maximal frustration. First, low-energy modes around the $M$ points can be thermally populated and contribute to the finite momentum-space integration windows used to extract the valley populations. A simple way to view this effect is to write the counted signal as a coherent three-mode component plus a weakly correlated background. If the coherent component contributes a fraction $f_c$ of the counted atoms, the contrast of the correlated three-mode fluctuation is reduced roughly by a factor $f_c^2$, even when the underlying bias-dependent structure of the condensate fluctuations is preserved. Consistent with this interpretation, reducing the momentum-space integration window increases the measured peak amplitude, indicating that background modes contribute to the reduction. However, this check does not fully recover the ideal three-mode value, so background dilution is unlikely to be the only limitation.

Second, the experiment averages over an inhomogeneous atomic cloud. Different spatial regions can have different local atom numbers and hence different effective interaction scales and relaxation histories. Such spatial averaging can smear the ideal susceptibility peak, especially near $\Delta E\simeq0$, where the response is most sensitive to small variations of local parameters.

Finally, the nearly degenerate point may be particularly sensitive to finite-time relaxation. The system is prepared from an initially $M_1$-dominated state and then evolves dissipatively toward the long-time measured state. Near $\Delta E\simeq0$, the free-energy landscape is exceptionally soft, and the relevant collective relaxation can be slower than away from the symmetric point, giving a critical-slowing-like limitation within the finite experimental evolution time of $130.1~\mathrm{ms}$. The experimentally observed hierarchy $\Delta n_1<\Delta n_2\simeq\Delta n_3$ around $\Delta E\simeq0$ may therefore indicate weak residual deviations from the ideal $C_3$-symmetric three-valley distribution, rather than contradicting the discrete three-mode interpretation.

These effects are not included in the ideal canonical calculation of the homogeneous discrete three-mode system. The three-mode model therefore captures the position and overall structure of the susceptibility peak, while the absolute amplitude can be reduced by thermally populated modes, spatial averaging and finite-time relaxation beyond the idealized limit.

\section*{Supplemental Note 4: Phase rigidity and harmonic-oscillator description}

This note supports the positive-bias analysis in the main text. We use the same two-valley conjugate variables introduced in Supplemental Note~2 and show how, once the $M_1$ valley is depleted, the pair-tunneling interaction produces a phase-rigid minimum rather than a barrier-softened landscape. Expanding around this minimum gives the macroscopic harmonic-oscillator description used to interpret Fig.~4.

\subsection*{S4A. Phase locking in the selected two-valley stripe phase}

For $\Delta E>0$, the energy bias suppresses the population of $M_1$ and selects the two-valley stripe phase formed by $M_2$ and $M_3$. The residual dynamics is governed by the relative phase and population imbalance of these two valleys, with $[\hat{\phi},\hat{\eta}]=2i$. In contrast to the negative-bias case, there is no macroscopically occupied $M_1$ reservoir anchoring the relative phase at $\phi=0$ or $\pi$.

The relevant two-valley interaction energy is
\begin{equation}
F_{23}
\simeq
\frac{U_1}{2}(N_2^2+N_3^2)
+
2U_2N_2N_3
+
U_2N_2N_3\cos(2\phi).
\label{eq:S4_F23}
\end{equation}
Since $U_2>0$, the phase-dependent term is minimized, whenever $N_2N_3$ is nonzero, by $\cos(2\phi)=-1$, corresponding to $\phi=\pm\pi/2$. The value of $N_2N_3$ controls the phase stiffness, but not the location of the phase minimum. The pair-tunneling interaction therefore locks the relative phase of the selected stripe phase.

At the phase-locked minimum, the interaction energy becomes
\begin{equation}
F_{23}(\phi=\pm\pi/2)\simeq\frac{U_1}{2}(N_2^2+N_3^2)+U_2N_2N_3=
\frac{U_1+U_2}{4}N_{\mathrm{act}}^2+\frac{U_1-U_2}{4}\eta^2 .
\label{eq:S4_F23_locked}
\end{equation}
For $U_1>U_2$, this energy confines the population imbalance around $\eta=0$. Thus phase locking at $\phi=\pm\pi/2$ and suppression of relative-population fluctuations arise from the same pair-tunneling interaction. This phase-rigid minimum is the starting point for the harmonic expansion below.

\subsection*{S4B. Harmonic expansion around the phase-locked minimum}

We now expand the two-valley energy around either phase-locked minimum, $\phi_0=\pm\pi/2$. We define the small phase fluctuation $\hat q=\hat{\phi}-\phi_0$ and the conjugate momentum $\hat p=\hat{\eta}/2$, so that $[\hat q,\hat p]=i$. Near the minimum, $\cos(2\hat{\phi})\simeq -1+2\hat q^2$. Keeping terms to quadratic order in $\hat q$ and $\hat{\eta}$ for fixed $N_{\mathrm{act}}$, and neglecting the higher-order coupling $\hat{\eta}^2\hat q^2$, the energy measured from the minimum is
\begin{equation}
\delta \hat F_{23}
\simeq
\frac{U_1-U_2}{4}\hat{\eta}^2
+
\frac{1}{2}U_2N_{\mathrm{act}}^2\hat q^2 .
\label{eq:S4_quadratic_eta}
\end{equation}
This is equivalently written as a harmonic oscillator,
\begin{equation}
\delta \hat F_{23}
\simeq
\frac{\hat p^2}{2m_{\mathrm{eff}}}
+
\frac{1}{2}K_q\hat q^2 ,
\label{eq:S4_QHO}
\end{equation}
with $m_{\mathrm{eff}}=1/[2(U_1-U_2)]$ and $K_q=U_2N_{\mathrm{act}}^2$. The corresponding oscillator energy scale is
\begin{equation}
\hbar\omega_p
=
\sqrt{\frac{K_q}{m_{\mathrm{eff}}}}
=
N_{\mathrm{act}}\sqrt{2U_2(U_1-U_2)} .
\label{eq:S4_plasma_frequency}
\end{equation}

Thus, within the harmonic expansion, the selected stripe phase is mapped onto a collective harmonic oscillator formed by the relative phase and valley-population imbalance. For the parameter set of Supplemental Note~1 and $N_{\mathrm{act}}\simeq N$ in the positive-bias stripe regime, Eq.~\eqref{eq:S4_plasma_frequency} gives a representative oscillator scale $\hbar\omega_p\simeq2.33~\mathrm{nK}$.

\subsection*{S4C. Quantum--thermal variance of the population imbalance}

The harmonic oscillator in Eq.~\eqref{eq:S4_QHO} gives the equilibrium variance of the conjugate momentum,
\[
\langle(\delta\hat p)^2\rangle
=
\frac{m_{\mathrm{eff}}\hbar\omega_p}{2}
\coth\left(
\frac{\hbar\omega_p}{2k_{\mathrm B}T}
\right).
\]
Because $\hat{\eta}=2\hat p$ and the selected stripe phase is symmetric under $M_2\leftrightarrow M_3$, one has $\langle\hat{\eta}\rangle=0$ and therefore
\begin{equation}
\langle(\delta\hat{\eta})^2\rangle
=
2m_{\mathrm{eff}}\hbar\omega_p
\coth\left(
\frac{\hbar\omega_p}{2k_{\mathrm B}T}
\right)
=
N_{\mathrm{act}}
\sqrt{\frac{2U_2}{U_1-U_2}}
\coth\left(
\frac{\hbar\omega_p}{2k_{\mathrm B}T}
\right).
\label{eq:S4_eta_variance}
\end{equation}

Equation~\eqref{eq:S4_eta_variance} connects the zero-point and thermally occupied oscillator regimes. In the quantum limit, $k_{\mathrm B}T\ll\hbar\omega_p$, it reduces to
\[
\langle(\delta\hat{\eta})^2\rangle
\simeq
N_{\mathrm{act}}
\sqrt{\frac{2U_2}{U_1-U_2}},
\]
where the residual population imbalance is set by zero-point motion. In the thermal limit, $k_{\mathrm B}T\gg\hbar\omega_p$, it becomes
\[
\langle(\delta\hat{\eta})^2\rangle
\simeq
\frac{2k_{\mathrm B}T}{U_1-U_2},
\]
which is the equipartition result for the quadratic imbalance energy in Eq.~\eqref{eq:S4_quadratic_eta}.

With the usual thermodynamic scaling of the interaction coefficients at fixed density, $U_{1,2}\propto1/N_{\mathrm{act}}$, the zero-point variance is extensive in $N_{\mathrm{act}}$. The thermal-limit expression has the same extensive scaling when the temperature is treated as an intensive physical scale. For the finite-temperature comparison in Fig.~4, we evaluate the full expression in Eq.~\eqref{eq:S4_eta_variance}. This collective-oscillator scaling underlies the suppression of the normalized relative variance, $\sigma_{\mathrm{rel}}^2\simeq\langle(\delta\hat{\eta})^2\rangle/N_{\mathrm{act}}^2$, in the phase-rigid stripe phase.

\subsection*{S4D. Comparison with canonical three-mode calculations}

The experimentally measured normalized relative variance is
\begin{equation}
\sigma_{\mathrm{rel}}^2
=
\frac{\langle(\delta n_2-\delta n_3)^2\rangle}
{\langle n_2+n_3\rangle^2}.
\label{eq:S4_sigma_def}
\end{equation}
In the selected two-valley stripe phase, this quantity corresponds to the imbalance variance normalized by the active population,
\begin{equation}
\sigma_{\mathrm{rel}}^2
\simeq
\frac{\langle(\delta\hat{\eta})^2\rangle}{N_{\mathrm{act}}^2},
\label{eq:S4_sigma_eta}
\end{equation}
where $N_{\mathrm{act}}=\langle\hat N_2+\hat N_3\rangle$. Combining Eq.~\eqref{eq:S4_sigma_eta} with Eq.~\eqref{eq:S4_eta_variance} gives
\begin{equation}
\sigma_{\mathrm{rel}}^2
=
\frac{1}{N_{\mathrm{act}}}
\sqrt{\frac{2U_2}{U_1-U_2}}
\coth\left(
\frac{\hbar\omega_p}{2k_{\mathrm B}T}
\right).
\label{eq:S4_sigma_HO}
\end{equation}

Equation~\eqref{eq:S4_sigma_HO} shows that the unnormalized imbalance variance is an oscillator fluctuation, while the normalized relative variance is suppressed by the active two-valley population. In the phase-rigid regime, the residual dynamics therefore resembles a coherent collective oscillator rather than sampling between macroscopically separated population configurations.

In Fig.~4(c) of the main text, the blue circles are the canonical three-mode calculations of $\sigma_{\mathrm{rel}}^2$ at $\Delta E=3~\mathrm{nK}$, where the population of $M_1$ is strongly suppressed. The orange curve is a fit to the canonical data using the harmonic-oscillator form of Eq.~\eqref{eq:S4_sigma_HO}. The fitted scales are consistent with the parameter estimates entering Eq.~\eqref{eq:S4_sigma_HO}, showing that the selected stripe phase is well described by the harmonic expansion around the phase-locked minimum. The vertical dash-dotted line marks the experimentally constrained value $T/N=0.04~\mathrm{nK}$ used for the main theoretical curves. The Husimi $Q$ distribution shown in the inset provides the corresponding phase-space visualization: it is localized near $\eta=0$ and $2\phi=\pm\pi$, illustrating phase rigidity and confinement of the conjugate population imbalance.

This comparison completes the positive-bias counterpart of the negative-bias phase-scrambling mechanism. In the negative-bias regime, number--phase conjugacy softens a phase-locked classical barrier and allows residual thermal fluctuations to sample competing minority-valley configurations. In the selected stripe phase, the same conjugate structure appears as phase rigidity: pair tunneling locks the relative phase, and the residual imbalance fluctuations are those of a collective oscillator interpolating between thermal equipartition and the quantum zero-point limit.

\end{document}